\documentclass[12pt]{revtex4}
\usepackage{amssymb}
\usepackage{latexsym}
\usepackage{epsfig}
\begin{document}                             

\title{Emergence of warm inflation in curved space-time between  accelerating branes}

\author{Aroonkumar Beesham\footnote{abeesham@yahoo.com}}
\affiliation{Department of Mathematical Sciences, University of Zululand, Private Bag X1001, Kwa-Dlangezwa 3886, South Africa}

\begin{abstract}
It appears that having  our own brane to somehow interact with other branes could give rise to quite an interesting system and that that interaction
could lead to some observable effects. We consider the
question of whether or not these signatures of interaction between the  branes can be observed. To answer  this question, we investigate the effect induced by the inflaton in the WMAP7 data using the warm  inflationary model. In this model, slow-roll and
perturbation parameters are
given in terms of the inflaton thermal distribution. We show that this distribution depends on  the orbital radius of the brane motion under the interaction potential of other branes in extra dimensions. Thus, an enhancement in the
brane inflation can be a signature of an orbital motion in extra dimensions and consequently,
some signals of other branes can be detected by observational data. According to experimental data, the $N\simeq 50$ case leads to $n_{s}\simeq 0.96$, where \emph{N} and $n_{s}$ are the number of e-folds and the spectral index, respectively. This standard case may be found in the range $0.01 < R_{Tensor-scalar } < 0.22$, where $R_{Tensor-scalar }$ is the tensor-scalar ratio. We find that at this point, the radial distance between our brane and another brane is $R=(1.5 GeV)^{-1}$ in intermediate, and $R=(0.02225 GeV)^{-1}$ in logamediate inflation.

\end{abstract}

\maketitle
\section{Introduction}
Recently, it was argued that  the boundary conditions to be imposed on the quantum state of the whole multiverse could be such that  brane universes could be created in entangled pairs \cite{m1}. Also, the consideration of entanglement between
the quantum states of two or more brane universes in a multiverse scenario provides
us with a completely new paradigm that opens the door to novel approaches for
traditionally unsolved problems in cosmology, more precisely, the problems of
the cosmological constant, the arrow of time and the choice of boundary conditions,
amongst others \cite{m2}. Some authors have tried to find direct evidence of the existence of other brane universes using a dark energy model \cite{m3}. Also some researchers show that other branes are made observable for us
through interaction with our own brane
\cite{m4}. In their paper, the orbital radius of our brane in extra dimensions can be described according to the interaction potential of other branes. In some scenarios, the properties of the interaction potential are calculated for a composite quantum state of two    branes whose states are quantum mechanically correlated \cite{m1,m2}. It appears that having our own brane to somehow interact with other branes could give rise to quite an interesting system, and that that interaction could lead to an orbital motion  in extra dimensions.

The main question is the possibility of considering the properties of  other branes against  observational data? The warm  inflationary model helps us to perform precision tests of the universal extra dimensional models, and explore the new physics
against observational data. In this scenario, after the  period of inflation, the radiation of the universe becomes dominant and the reheating epoch will not happen. The results of this model are compatible with WMAP7 and Planck data \cite{m5}. In this theory, slow-roll and
perturbation parameters are
given in terms of  the thermal distribution of the inflaton. On the other hand, this distribution is
given in terms of the orbital radius of the brane motion \cite{m4} in extra dimensions. As the  interaction potential increases, the effect of the inflaton radiation from the 
horizon that appears in the brane-antibrane system on the universe's inflation becomes systematically
more effective because at higher energies, there exist more channels for inflaton production and its decay into particles.

   The outline of the paper is as the following. In section \ref{o1}, we
consider the effect of the orbital radius of the brane motion under the interaction potential of the other branes on the thermal distribution of inflatons. In section \ref{o2}, using the warm  inflationary model, we
analyze the signature of other branes against observational data. The last section is devoted to a summary and conclusion.

\section{  The thermal distribution of inflatons near the appeared horizon in the brane-antibrane system}\label{o1}
Previously, the dynamical behavior of a pair of Dp and anti Dp branes
which move parallel to each other in the region that the brane and antibrane annihilation
will not occur was considered \cite{m4}. Also, the orbital radius of the brane motion due to the interaction potential in extra dimensions was studied. Using these results, we calculate the thermal distribution of inflatons near the horizon that appears in the brane-antibrane system and show that the thermal distribution of inflatons can be given in terms of the orbital radius of the brane motion in extra dimensions.

The d-dimensional metric in the brane-antibrane system is expressed as:
   \begin{eqnarray}
ds^{2}=g_{\mu\nu}dx^{\mu}dx^{\nu}+g_{\rho\sigma}dx^{\rho}dx^{\sigma}+g_{ab}dx^{a}dx^{b}
\label{w1}
\end{eqnarray}
where $g_{\mu\nu}$ and $g_{\rho\sigma}$ are the p-dimensional metrics along the Dp   and the anti Dp brane, respectively, and $g_{ab}$ is the (d-2p) dimensional metric along the transverse coordinates.

Now let us consider the wave equation of the inflaton in extra dimensions between two branes:
\begin{equation}\label{w2}
\left\{-\frac{\partial^{2}}{c^{2}\partial \chi^{2}}+\frac{\partial^{2}}{\partial r^{2}}\right\}B=0
\end{equation}
where $\chi, r$ are the transverse coordinates between the two branes. This equation  corresponds to flat space-time. The interaction potential between the Dp brane-anti Dp brane in extra dimensions is of the type  \cite{m4}
\begin{eqnarray}
V(R)\sim\frac{64\pi^{2}\mu^{4}}{27}\label{w3}
\end{eqnarray}
where $\mu^{4}=\frac{27}{32\pi^{2}}T_{3}h^{4}$, $ h(R)=\frac{b^{4}}{R^{4}}$, \emph{R} is the orbital radius distance between the two branes, $T_{3}$  the brane tension and \emph{b} the curvature radius of the $AdS_{5}$ throat. This potential leads to curved space-time.

Thus, to write the inflaton wave equation in curved space-time, we should use the following reparameterizations:
 \begin{eqnarray}
&&r\rightarrow \rho(r,\chi)\nonumber \\  &&\chi\rightarrow \tau(r,\chi)
\label{w4}
\end{eqnarray}
that leads to the following inflaton wave equation:
 \begin{eqnarray}
&&[\left\{(\frac{\partial \tau}{\partial r})^{2}-(\frac{\partial \tau}{\partial \chi})^{2}\right\}\frac{\partial^{2}}{c^{2}\partial\tau^{2}}\nonumber \\&& + \{(\frac{\partial \rho}{\partial r})^{2}-(\frac{\partial \rho}{\partial \chi})^{2}\}\frac{\partial^{2}}{\partial\rho^{2}}]B=0
\label{w5}
\end{eqnarray}
 We can normalize the distance between the two branes to unity by making the following choices:
 \begin{eqnarray}
&&\rho(r,\chi)=\frac{r}{R(\chi)}\nonumber \\&&\tau=\beta c^{2}\int_{0}^{\chi}d\acute{t}\frac{R(\acute{\chi})}{\dot{R}(\acute{\chi})}-\beta\frac{r^{2}}{2}
\label{w6}
\end{eqnarray}

         With the above considerations, the wave equation is written as:
        \begin{equation}\label{w7}
(-g)^{1/2}\frac{\partial}{\partial x^{\mu}}[g^{\mu\nu}(-g)^{1/2}\frac{\partial}{\partial x^{\nu}}]B=0
\end{equation}
 where $ x^{5}=\tau,x^{4}=\rho$ and the metric elements are obtained as:
  \begin{eqnarray}
&&g^{\tau\tau}=-\frac{1}{\beta^{2}c^{2}}(\frac{R}{\dot{R}})(\frac{1-\frac{\dot{R}^{2}}{c^{2}}\rho^{2}}{1+\frac{\dot{R}}{c^{2}}\rho^{2}})\nonumber \\ && g^{44}=R^{2}(\frac{1+\frac{\dot{R}^{2}}{c^{2}}\rho^{2}}{1-\frac{\dot{R}}{c^{2}}\rho^{2}})
\label{w8}
\end{eqnarray}
The horizon of this system is located at:
    \begin{equation}\label{w9}
r_{horizon}=\frac{c R}{\dot{R}}
\end{equation}
where \emph{c} is velocity of light. In Kruskal coordinates the metric of system becomes \cite{m6,m7}:
\begin{eqnarray}
&&ds^{2}=g_{\mu\nu}dx^{\mu}dx^{\nu}+g_{\rho\sigma}dx^{\rho}dx^{\sigma}-r_{horizon}\frac{e^{-\frac{r}{r_{horizon}}}}{r}d\bar{u}d\bar{v}+r^{2}d\theta^{2}\nonumber \\&&
\bar{u}=-2r_{horizon}e^{-u/2r_{horizon}},\bar{v}=-2r_{horizon}e^{-v/2r_{horizon}}\nonumber \\&&
u=\chi-r^{\ast},v=\chi+r^{\ast},r^{\ast}=-r-r_{horizon}\ln{|r-r_{horizon}|}
\label{w10}
\end{eqnarray}
  Since
the Killing vector in Kruskal coordinates is given by $ \frac{\partial}{\partial \bar{u}}$
on the past horizon $H^{-}$, the positive frequency normal mode solution in Kruskal coordinates is approximated by:
\begin{equation}\label{w11}
B \propto e^{-i\omega\bar{u}}
\end{equation}
where $\omega$ is the inflaton energy in extra dimensions. Using this fact that $\bar{v} = 0$ on $H^{-}$ \cite{m7} we can estimate the original
positive frequency normal mode on the past horizon as:
\begin{eqnarray}
B & \propto & e^{-i\omega u}=(\frac{|\bar{u}|}{2r_{horizon}})^{-i2r_{horizon}\omega}=
\{
\begin{array}{cc}
(-\frac{\bar{u}}{2r_{horizon}})^{-i2r_{horizon}\omega} & \text(region I) \\
(\frac{\bar{u}}{2r_{horizon}})^{-i2r_{horizon}r_{horizon}} & \text(region II)
\end{array}\label{w12}
\end{eqnarray}

In Eq. (\ref{w12}), we can use the fact that
 $(-1)^{-i2r_{horizon}\omega}=e^{2r_{horizon}\omega}$. Using equation (\ref{w12}) we observe that the inflaton states in the
horizon satisfy the following condition \cite{m7,m8}:
\begin{eqnarray}
&&(B_{out}-\tanh r_{\omega}B_{in})|\text{system}\rangle_{in\bigotimes out}=0 \nonumber \\&& \tanh r_{\omega}=e^{-2r_{horizon}\omega}\label{w13}
\end{eqnarray}
which actually constitutes a boundary state. In fact, we can view Hawking
radiation as the pair creation of a positive energy field that goes to infinity and a
negative energy field that falls into the horizon of the brane-antibrane system. The pair is created in a
particular entangled state. So the Unruh state can be viewed as an entangled
thermal state. The above definition of the positive frequency solution in terms of $B_{out} $ and $B_{in} $ leads to the Bogoliubov transformation \cite{m6,m7,m8} for the particle creation and annihilation operators in the brane-antibrane system and Minkowski space-times in the exterior region of the system:
\begin{eqnarray}
&&d=coshr_{\omega}\alpha_{out}-sinhr_{\omega}\alpha^{\dag}_{in} \nonumber \\&&
d^{\dag}=coshr_{\omega}\alpha^{\dag}_{out}-sinhr_{\omega}\alpha_{in} \nonumber \\&&
tanhr_{\omega}=e^{-2\pi r_{horizon}\omega}
\label{w14}
\end{eqnarray}
where $ d^{\dag}$ and \emph{\emph{d}} are the creation and annihilation operators, respectively, acting on the Minkowski vacuum, $ \alpha^{\dag}_{out}$ and $ \alpha_{out}$  the respective operators acting on the brane-antibrane vacuum outside the event horizon, and $ \alpha^{\dag}_{in}$ and $ \alpha_{in}$ are the respective  operators acting on the brane-antibrane vacuum inside the event horizon. 

Thus, we can write the Bogoliubov transformation between the Minkowski and curved creation and annihilation operators as:
\begin{equation}\label{w15}
d|\text{system}\rangle_{out\bigotimes in}=(\alpha_{out}-\tanh r_{\omega}\alpha_{in}^{\dag})|\text{system}\rangle_{out\bigotimes in}=0
\end{equation}
which actually constitutes a boundary state.
Now, we assume that the system vacuum $|\text{system}\rangle_{out\bigotimes in}$ is related to the flat vacuum $|0\rangle_{flat}$ by
\begin{equation}\label{w16}
|\text{system}\rangle_{out\bigotimes in}=F|0\rangle_{flat}
\end{equation}
where $F$ is a function to be determined later.

From $[\alpha_{out},\alpha_{out}^{\dag}]=1$, we obtain $[\alpha_{out},(\alpha_{out}^{\dag})^{m}]=\frac{\partial}{\partial \alpha_{out}^{\dag} }(\alpha_{out}^{\dag})^{m}$ and $[\alpha_{out},F]=\frac{\partial}{\partial \alpha_{out}^{\dag} }F$. Then using Eqs.(\ref{w15}) and (\ref{w16}), we get the following differential equations for $F$:
\begin{equation}\label{w17}
 (\frac{\partial F}{\partial \alpha_{out}^{\dag}}-tanh r_{\omega}\alpha_{in}^{\dag}F)=0
\end{equation}
and the solution is given by
\begin{equation}\label{w18}
 F_=e^{tanhr_{\omega}\alpha_{out}^{\dag}\alpha_{in}^{\dag}}
\end{equation}
By substituting Eq. (\ref{w18}) into Eq. (\ref{w16}) and by properly normalizing the state vector, we get
\begin{eqnarray}&&|\text{system}\rangle_{out\bigotimes in}
=  N e^{tanhr_{\omega}\alpha_{in}^{\dag}\alpha_{out}^{\dag}} |0\rangle_{flat} \nonumber \\&&
= \frac{1}{coshr_{\omega}} \sum_{m}tanh^{m}r_{\omega}|m\rangle_{out}\otimes|\bar{m}\rangle_{in}\label{w19}
\end{eqnarray}
where $|m\rangle_{in}$ and $|\bar{m}\rangle_{out}$ are the orthonormal bases (normal mode solutions) for a particle that acts on $H_{in}$ and $H_{out}$ respectively, and $N$ is the normalization constant.

       Eq. (\ref{w19}) expresses that the states inside and outside the horizon are entangled. However, this entanglement depends on the event horizon and the horizon is given in terms of \emph{R}, the orbital radius of the brane motion in the intraction potential of the other brane, $ r_{horizon}=\frac{cR}{\dot{R}}$, and consequently, the entanglement changes with the orbital radial distance between the two branes.
       We derive the thermal distribution for inflatons in extra dimensions as the following:
 \begin{eqnarray}
 &&<B>= _{out\bigotimes in}\langle \text{system}|\alpha_{in}^{\dag}  \alpha_{in}|\text{system}\rangle_{out\bigotimes in} \nonumber\\ && = \frac{e^{-2\pi r_{horizon} \omega}}{1-e^{-2\pi  r_{horizon} \omega}}  \label{w20}
\end{eqnarray}
The above equation shows that different numbers of inflatons are produced with different probabilities inside and outside of the apparent horizon in the brane-antibrane system. These probabilities are related to the orbital radial distance of the two branes and the energy of the inflatons.

\section{Considering the effect of other branes on cosmic inflation by using warm  inflationary model
 }\label{o2}
In this section we enter the effects of the interaction potential between the branes on the results of the derivation of slow-roll,
perturbation parameters and other important parameters in the inflationary model \cite{m5}. We show that these parameters are given in terms of the orbital radial distance between the two branes and describe
the shape of the interaction potential between branes. Also, using the  inflationary model, we discuss the signature of interaction between branes against observational data.

 Previously, it has been shown that in the FRW brane with the metric
 \begin{eqnarray}
ds^{2}=g_{\mu\nu}dx^{\mu}dx^{\nu}=-dt^{2}-a^{2}(t)dx^{i}dx_{i}
\label{w21}
\end{eqnarray}
the dynamics of warm  inflation is presented by these equations \cite{m5}:
 \begin{eqnarray}
  &&\dot{\rho}+3H(P+\rho)=-\Gamma <\dot{B}>^{2}\nonumber \\
&&\dot{\rho}_{\gamma}+4H\rho_{\gamma}=-\Gamma <\dot{B}>^{2}\nonumber \\
&&H^{2}=\frac{1}{2}(<\dot{B}>^{2}+V(B))+\frac{1}{3}\rho_{\gamma}\nonumber \\
&&V(B)=m^{2}<B>^{2}
\label{w22}
\end{eqnarray}
 where $\rho$ is the energy density, \emph{p}  the pressure, $\rho_{\gamma}$
 the energy density of the radiation, $\Gamma$  the dissipative coefficient, $<B>$  the thermal distribution of the inflaton, and the overdot ($\dot{}$) is the derivative with respect to  cosmic time. In the previous section, we discussed that the thermal distribution of the  inflaton can be given as a function of the orbital radial distance between branes. Using this fact, we can rewrite the above equation as:
 \begin{eqnarray}
  &&\dot{\rho}+3H(P+\rho)=-\Gamma (\frac{\ddot{R}R-\dot{R}}{2\pi \omega R^{2}})^{2}\nonumber \\&&
\dot{\rho}_{\gamma}+4H\rho_{\gamma}=-\Gamma (\frac{\ddot{R}R-\dot{R}}{2\pi \omega R^{2}})^{2}\nonumber \\&&
H^{2}=\frac{1}{2}((\frac{\ddot{R}R-\dot{R}}{2\pi \omega R^{2}})^{2}+V(R,\dot{R}))+\frac{1}{3}\rho_{\gamma}\nonumber \\&&
V(R,\dot{R})=m^{2}(1-\frac{\dot{R}}{2\pi \omega R})^{2}
\label{w23}
\end{eqnarray}
Using quantum field theory
methods \cite{m9,m10}, the dissipation coefficient ($\Gamma$) in the above equations could be calculated as:
\begin{eqnarray}
\Gamma=\Gamma_{0}\frac{T^{3}}{<B>^{2}}\sim \Gamma_{0}\frac{4\pi^{2}\omega^{2}T^{3}R^{2}}{\dot{R}^{2}}
\label{w24}
\end{eqnarray}
where
T is the temperature of the thermal bath. During the inflationary epoch, the energy density $\rho$ is more than the radiation energy density
$\rho>\rho_{\gamma}$; however, it is comparable with the potential energy density $V (B^{2})$ ($\rho\sim V$ ) \cite{m5}. The slow-roll approximation ($<\ddot{B}>\leq(3H+\frac{\Gamma}{3})<\dot{B}>$) \cite{m11} with the condition that inflation
radiation production be quasi-stable, ($\dot{\rho}_{\gamma}\leq 4H\rho_{\gamma},\dot{\rho}_{\gamma}\leq \Gamma <\dot{B}>$ ) leads to following dynamic equations \cite{m5}
 \begin{eqnarray}
  &&3H(1+\frac{r}{3})<\dot{B}>=-\frac{1}{2}\acute{V}\nonumber \\
&&\rho_{\gamma}=\frac{3}{4}r<\dot{B}>^{2}=\frac{r}{(1+\frac{r}{3})^{2}}\frac{\acute{V}^{2}}{V}=CT^{4}\nonumber \\&&
H^{2}=\frac{1}{2}V
\label{w25}
\end{eqnarray}
where $r=\frac{\Gamma}{3H}$, and $C=\frac{\pi^{2}g^{\ast}}{30}$
($g^{\ast}$ are the number of relativistic degrees of freedom). In the
above equations, a prime ($\acute{}$) denotes a derivative with respect to the field \textbf{B}. Using this equation and the thermal distribution of the inflaton in Eq. (\ref{w20}), we can obtain the dynamic equations with respect to \emph{R}, the orbital radial distance between the two branes:
\begin{eqnarray}
  &&3H(1+\frac{r}{3})(\frac{\ddot{R}R-\dot{R}}{2\pi \omega R^{2}})=-\frac{1}{2}\acute{V}(R,\dot{R})\nonumber \\&&
\rho_{\gamma}=\frac{3}{4}r(\frac{\ddot{R}R-\dot{R}}{2\pi \omega R^{2}})^{2}=\frac{r}{(1+\frac{r}{3})^{2}}\frac{\acute{V}^{2}(R,\dot{R})}{V(R,\dot{R})}=CT^{4}\nonumber \\&&
H^{2}=\frac{1}{2}V(R,\dot{R})
\label{w26}
\end{eqnarray}
where the prime ($\acute{}$) denotes derivative with respect to  \emph{R}. From the above equations, 
the temperature of the thermal bath is given by\cite{m5}:
\begin{eqnarray}
T=[-\frac{r\dot{H}}{2C(1+\frac{r}{3})}]^{\frac{1}{4}}=
[-\frac{r(\ddot{R}R-\dot{R})}{4C\pi \omega R^{2}(1+\frac{r}{3})}]^{\frac{1}{4}}
\label{w27}
\end{eqnarray}
This temperature depends on the orbital radial distance between the two branes. As the branes come close to each other, the temperature of the thermal bath increases. The reason for this is as follows: with decreasing distance between the two branes, the interaction potential increases and more inflatons radiate from the apparent horizon of the brane-antibrane system.

At this stage, we tend to calculate the dependency of slow-roll parameters on the orbital radial distance between different branes. These parameters in warm  inflation are \cite{m5}:
 \begin{eqnarray}
  &&\epsilon=-\frac{1}{H}\frac{d}{dt}\ln(H)\nonumber \\&&
\eta=-\frac{\ddot{H}}{H\dot{H}}
\label{w28}
\end{eqnarray}
where $H=\frac{\dot{a}}{a}$ and \emph{a} is the scale factor. To calculate these parameters, we should determine the explicit form of the scale factor.

Until now, eight possible asymptotic solutions for cosmological dynamics have been proposed \cite{m12}. Three of these solutions have non-inflationary scale factor and another three  solutions give de Sitter, intermediate and power-low
inflationary expansion. Finally, two cases of these solutions have asymptotic expansion with scale factor $(a = a_{0} exp(A(ln t)^{\lambda})$. This version of inflation is named logamediate inflation \cite{m13}. In this paper, we will study the warm-tachyon inflationary model in the scenarios of intermediate  and logamediate inflation.

 Firstly, let us  consider intermediate inflationary expansion. In this model, the expansion of the universe
is between standard de Sitter inflation with scale factor $a(t) = a_{0} exp(H_{0}t)$ and power law inflation with scale factor $a(t) = t^{p}, p > 1$ (slower than the first one) \cite{m14,m15}. The scale factor of this model
has the  form below \cite{m16,m17}:
\begin{eqnarray}
a=a_{0}exp(At^{f}),0<f<1
\label{w29}
\end{eqnarray}
where \emph{A} is a positive constant.  The number of e-folds in this case is \cite{m5}:
\begin{eqnarray}
N=\int_{t_{1}}^{t}H dt=A(t^{f}-t_{1}^{f})
\label{w30}
\end{eqnarray}
where $t_{1}$ is the begining time of inflation.
From Eqs.(\ref{w20}), (\ref{w24}), (\ref{w25}), (\ref{w26}), (\ref{w27}) and (\ref{w29}) we obtain the Hubble parameter as:
 \begin{eqnarray}
 && H=fA(\frac{\ln <B>-\ln <B_{0}>}{\bar{\omega}})^{\frac{8(f-1)}{5f+2}}=\nonumber \\&&
fA(\frac{\ln \frac{e^{-2\pi r_{horizon} \omega}}{1-e^{-2\pi  r_{horizon} \omega}}-\ln \frac{e^{-2\pi r_{0,horizon} \omega}}{1-e^{-2\pi  r_{0,horizon} \omega}}}{\bar{\omega}})^{\frac{8(f-1)}{5f+2}}\sim\nonumber \\&&
fA(\frac{-2\pi \omega(r_{horizon}-r_{0,horizon})+\ln \frac{1-e^{-2\pi  r_{0,horizon} \omega}}{1-e^{-2\pi  r_{horizon} \omega}}}{\bar{\omega}})^{\frac{8(f-1)}{5f+2}}\sim\nonumber \\&& fA(\frac{2\pi \omega(\frac{R_{0}}{\dot{R}_{0}}-\frac{R}{\dot{R}})+\ln \frac{1-e^{-2\pi  \frac{R_{0}}{\dot{R}_{0}} \omega}}{1-e^{-2\pi  \frac{R}{\dot{R}} \omega}}}{\bar{\omega}})^{\frac{8(f-1)}{5f+2}}\nonumber \\&& B=B_{0}exp(\bar{\omega}t^{\frac{5f+2}{8}})
\label{w31}
\end{eqnarray}
where $ \bar{\omega}=(\frac{6}{\Gamma_{0}}(\frac{2C}{3})^{\frac{3}{4}})^{\frac{1}{2}}(\frac{8(fA)^{\frac{5}{8}}(1-f)^{\frac{1}{8}}}{5f+2})$ and $\Gamma_{0}=constant$. This equation insists that the evolution of our brane universe  is affected by the number of inflatons that are radiated from the apparent horizon of the  brane-antibrane system and it changes with an increase or decrease in the orbital radial  distance between the two branes.

 The important slow-roll parameters $\epsilon$ and
$\eta$ are given  by:
 \begin{eqnarray}
 &&\epsilon=\frac{1-f}{fA}(\frac{\ln <B>-\ln <B_{0}>}{\bar{\omega}})^{-\frac{8f}{5f+2}}=\nonumber \\&&
\frac{1-f}{fA}(\frac{\ln \frac{e^{-2\pi r_{horizon} \omega}}{1-e^{-2\pi  r_{horizon} \omega}}-\ln \frac{e^{-2\pi r_{0,horizon} \omega}}{1-e^{-2\pi  r_{0,horizon} \omega}}}{\bar{\omega}})^{-\frac{8f}{5f+2}}\sim\nonumber \\&&
\frac{1-f}{fA}(\frac{-2\pi \omega(r_{horizon}-r_{0,horizon})+\ln \frac{1-e^{-2\pi  r_{0,horizon} \omega}}{1-e^{-2\pi  r_{horizon} \omega}}}{\bar{\omega}})^{-\frac{8f}{5f+2}}\sim\nonumber \\&&\frac{1-f}{fA}(\frac{2\pi \omega(\frac{R_{0}}{\dot{R}_{0}}-\frac{R}{\dot{R}})+\ln \frac{1-e^{-2\pi  \frac{R_{0}}{\dot{R}_{0}} \omega}}{1-e^{-2\pi  \frac{R}{\dot{R}} \omega}}}{\bar{\omega}})^{-\frac{8f}{5f+2}}
\label{w32}
\end{eqnarray}
and
\begin{eqnarray}
 &&\eta=\frac{2-f}{fA}(\frac{\ln <B>-\ln <B_{0}>}{\bar{\omega}})^{-\frac{8f}{5f+2}}=\nonumber \\&&
\frac{2-f}{fA}(\frac{\ln \frac{e^{-2\pi r_{horizon} \omega}}{1-e^{-2\pi  r_{horizon} \omega}}-\ln \frac{e^{-2\pi r_{0,horizon} \omega}}{1-e^{-2\pi  r_{0,horizon} \omega}}}{\bar{\omega}})^{-\frac{8f}{5f+2}}\sim\nonumber \\&&
\frac{2-f}{fA}(\frac{-2\pi \omega(r_{horizon}-r_{0,horizon})+\ln \frac{1-e^{-2\pi  r_{0,horizon} \omega}}{1-e^{-2\pi  r_{horizon} \omega}}}{\bar{\omega}})\sim\nonumber \\&&\frac{2-f}{fA}(\frac{2\pi \omega(\frac{R_{0}}{\dot{R}_{0}}-\frac{R}{\dot{R}})+\ln \frac{1-e^{-2\pi  \frac{R_{0}}{\dot{R}_{0}} \omega}}{1-e^{-2\pi  \frac{R}{\dot{R}} \omega}}}{\bar{\omega}})^{-\frac{8f}{5f+2}}
\label{w33}
\end{eqnarray}
respectively. These parameters depend on the orbital radial distance between the branes. With a decrease in this distance, more inflatons  are radiated from the apparent horizon of the system, the slow-roll parameters increase, and as a result, the universe inflates more.

 The energy density of radiation in this case has the following form:
\begin{eqnarray}
 &&\rho_{\gamma}=3(1-f)fA(\frac{\ln <B>-\ln <B_{0}>}{\bar{\omega}})^{\frac{8f-2}{5f+2}}=\nonumber \\&&
3(1-f)fA(\frac{\ln \frac{e^{-2\pi r_{horizon} \omega}}{1-e^{-2\pi  r_{horizon} \omega}}-\ln \frac{e^{-2\pi r_{0,horizon} \omega}}{1-e^{-2\pi  r_{0,horizon} \omega}}}{\bar{\omega}})^{\frac{8f-2}{5f+2}}\sim\nonumber \\&&
3(1-f)fA(\frac{-2\pi \omega(r_{horizon}-r_{0,horizon})+\ln \frac{1-e^{-2\pi  r_{0,horizon} \omega}}{1-e^{-2\pi  r_{horizon} \omega}}}{\bar{\omega}})^{\frac{8f-2}{5f+2}}
\sim\nonumber \\&&
3(1-f)fA(\frac{2\pi \omega(\frac{R_{0}}{\dot{R}_{0}}-\frac{R}{\dot{R}})+\ln \frac{1-e^{-2\pi  \frac{R_{0}}{\dot{R}_{0}} \omega}}{1-e^{-2\pi  \frac{R}{\dot{R}} \omega}}}{\bar{\omega}})^{\frac{8f-2}{5f+2}}
\label{w34}
\end{eqnarray}
According to this result, the radiation energy density is
given in terms of the orbital radius of the brane motion in extra dimensions. As the  interaction potential increases, the effect of the inflaton radiation from the apparent 
horizon in the brane-antibrane system on cosmic inflation becomes systematically
more effective because at higher energies there exist more channels for inflaton production.

Using Eqs. (\ref{w30}) and (\ref{w31}), the number of e-folds between the two fields $B_{1}$ and \textbf{B} is given by:
\begin{eqnarray}
&&N=A[(\frac{\ln <B>-\ln <B_{0}>}{\bar{\omega}})^{-\frac{8f}{5f+2}}-(\frac{\ln <B_{1}>-\ln <B_{0}>}{\bar{\omega}})^{-\frac{8f}{5f+2}}]=\nonumber \\&&
A[(\frac{\ln \frac{e^{-2\pi r_{horizon} \omega}}{1-e^{-2\pi  r_{horizon} \omega}}-\ln \frac{e^{-2\pi r_{0,horizon} \omega}}{1-e^{-2\pi  r_{0,horizon} \omega}}}{\bar{\omega}})^{-\frac{8f}{5f+2}}-\nonumber \\&&(\frac{\ln \frac{e^{-2\pi r_{1,horizon} \omega}}{1-e^{-2\pi  r_{1,horizon} \omega}}-\ln \frac{e^{-2\pi r_{0,horizon} \omega}}{1-e^{-2\pi  r_{0,horizon} \omega}}}{\bar{\omega}})^{-\frac{8f}{5f+2}}]\sim\nonumber \\&&
A[(\frac{-2\pi \omega(r_{horizon}-r_{0,horizon})+\ln \frac{1-e^{-2\pi  r_{0,horizon} \omega}}{1-e^{-2\pi  r_{horizon} \omega}}}{\bar{\omega}})^{-\frac{8f}{5f+2}}-\nonumber \\&&
(\frac{-2\pi \omega(r_{1,horizon}-r_{0,horizon})+\ln \frac{1-e^{-2\pi  r_{0,horizon} \omega}}{1-e^{-2\pi  r_{1,horizon} \omega}}}{\bar{\omega}})^{-\frac{8f}{5f+2}}]\sim\nonumber \\&&A[(\frac{2\pi \omega(\frac{R_{0}}{\dot{R}_{0}}-\frac{R}{\dot{R}})+\ln \frac{1-e^{-2\pi  \frac{R_{0}}{\dot{R}_{0}} \omega}}{1-e^{-2\pi  \frac{R}{\dot{R}} \omega}}}{\bar{\omega}})^{-\frac{8f}{5f+2}}-\nonumber \\&&
(\frac{2\pi \omega\{(\frac{R_{0}}{\dot{R}_{0}}-\frac{R_{1}}{\dot{R}_{1}})+\ln \frac{1-e^{-2\pi  \frac{R_{0}}{\dot{R}_{0}} \omega}}{1-e^{-2\pi  \frac{R_{1}}{\dot{R_{1}}} \omega}}}{\bar{\omega}})^{-\frac{8f}{5f+2}}]
\label{w35}
\end{eqnarray}
This equation depends on the $<B_{1}>$ and $<B_{0}>$. To obtain the explicit form of the number of e-folds in terms of the orbital radius distance between the branes, we should find the relation between $<B_{1}>$ and $<B_{0}>$ . At the begining of the inflation period where $\epsilon=1$, the inflaton in terms of constant
parameters of the model is:
\begin{eqnarray}
&&<B_{1}>=<B_{0}>exp(\bar{\omega}(\frac{1-f}{fA})^{\frac{5f+2}{8f}})\rightarrow\nonumber \\&&
\frac{e^{-2\pi r_{1,horizon} \omega}}{1-e^{-2\pi  r_{1,horizon} \omega}}=\frac{e^{-2\pi r_{0,horizon} \omega}}{1-e^{-2\pi  r_{0,horizon} \omega}}exp(\bar{\omega}(\frac{1-f}{fA})^{\frac{5f+2}{8f}})\rightarrow\nonumber \\&&
(r_{1,horizon})^{-1}\sim(r_{0,horizon})^{-1}exp(\bar{\omega}(\frac{1-f}{fA})^{\frac{5f+2}{8f}})-1\label{w36}
\end{eqnarray}
From the above equations, we obtain the inflaton (\textbf{B}(t)) and the distance between the two branes (R(t)) in terms of number of e-folds
\begin{eqnarray}
&&<B(t)>=<B_{0}>exp(\bar{\omega}(\frac{N}{A}+\frac{1-f}{fA})^{\frac{5f+2}{8f}})\rightarrow\nonumber \\&&
\frac{e^{-2\pi r_{horizon} \omega}}{1-e^{-2\pi  r_{horizon} \omega}}=\frac{e^{-2\pi r_{0,horizon} \omega}}{1-e^{-2\pi  r_{0,horizon} \omega}}exp(\bar{\omega}(\frac{N}{A}+\frac{1-f}{fA})^{\frac{5f+2}{8f}})\rightarrow\nonumber \\&&
(r_{horizon})^{-1}\sim(r_{0,horizon})^{-1}exp(\bar{\omega}(\frac{N}{A}+\frac{1-f}{fA})^{\frac{5f+2}{8f}})-1\rightarrow
\nonumber \\&&R(t)=R_{0}exp(-\int dt r_{horizon}(N,t))
\label{w37}
\end{eqnarray}
This equation shows that the orbital radial distance between the brane universes depends on the number of e-folds. This means that as the distance between the brane decreases,  more inflatons are created near the apparent horizon of the brane-antibrane system, and the number of e-folds increases.
\begin{figure}\epsfysize=10cm
{ \epsfbox{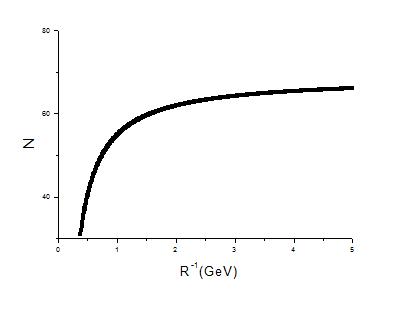}}\caption{The number of e-folds \emph{N} for intermediate
scenario as a function of the $R^{-1}$ for $R_{0}=0.45 (GeV)^{-1}$, $\omega=4.6 (GeV)$, $\dot{R}_{0}=0.01$, $\dot{R}=0.1$, \emph{A}=1 and $f=\frac{1}{2}$. } \label{1}
\end{figure}

In Fig.1 we present the number of e-folds \emph{N} for the intermediate
scenario as a function of  $R^{-1}$, where \emph{R} is the orbital radial distance between branes. In this plot, we choose $R_{0}=0.45 (GeV)^{-1}$, $\omega=4.6 (GeV)$, $\dot{R}_{0}=0.01$, $\dot{R}=0.1$, \emph{A}=1 and $f=\frac{1}{2}$. It is clear that the number of e-folds \emph{N} is much larger for a smaller orbital radial distance between the branes. This is because, as the distance between the branes becomes smaller, the temperature becomes larger and the thermal radiation of  the inflatons enhances.

Now, we will consider tensor and scalar perturbations that 
appear during the  inflationary period for the warm  inflation model. These perturbations may
leave an imprint in the CMB anisotropy and on the LSS \cite{m18,m19}. The power spectrum and a
spectral index are characteristics of each fluctuation: $\Delta_{R}^{2}(k)$ and $n_{s}$ for scalar perturbations, $\Delta_{T}^{2}(k)$ and $n_{T}$ for tensor perturbations. In warm and cool inflation models, the scalar power
spectrum is given by \cite{m5}:
\begin{eqnarray}
 \Delta_{R}^{2}=(\frac{H}{<\dot{B}>}<\delta B>)^{2}\label{w38}
\end{eqnarray}
where the thermal fluctuation in the warm inflation model yields \cite{m18,m19}:
\begin{eqnarray}
<\delta B>=(\frac{\Gamma H T^{2}}{(4\pi)^{3}})^{\frac{1}{4}}\label{w39}
\end{eqnarray}
Using Eqs. (\ref{w20}), (\ref{w37}), (\ref{w38}) and (\ref{w39}), we calculate the scalar power spectrum as:
\begin{eqnarray}
  &&\Delta_{R}^{2}=-(\frac{\Gamma_{0}^{3}}{36(4\pi)^{3}})^{\frac{1}{2}}\frac{H^{\frac{3}{2}}}{\dot{H}}=\nonumber \\&&(\frac{\Gamma_{0}^{3}}{36(4\pi)^{3}})^{\frac{1}{2}}(\frac{3^{11}(fA)^{15}(1-f)^{3}}{(2C)^{11}})^{\frac{1}{8}}
  <B>^{3}(\frac{\ln <B>-\ln <B_{0}>}{\bar{\omega}})^{-\frac{15f-18}{5f+2}}=\nonumber \\&&
  (\frac{\Gamma_{0}^{3}}{36(4\pi)^{3}})^{\frac{1}{2}}(\frac{3^{11}(fA)^{15}(1-f)^{3}}{(2C)^{11}})^{\frac{1}{8}}
  (\frac{e^{-2\pi r_{horizon} \omega}}{1-e^{-2\pi  r_{horizon} \omega}})^{3}\times\nonumber \\&&(\frac{\ln \frac{e^{-2\pi r_{horizon} \omega}}{1-e^{-2\pi  r_{horizon} \omega}}-\ln \frac{e^{-2\pi r_{0,horizon} \omega}}{1-e^{-2\pi  r_{0,horizon} \omega}}}{\bar{\omega}})^{-\frac{15f-18}{5f+2}}\sim\nonumber \\&&
 (\frac{\Gamma_{0}^{3}}{36(4\pi)^{3}})^{\frac{1}{2}}(\frac{3^{11}(fA)^{15}(1-f)^{3}}{(2C)^{11}})^{\frac{1}{8}}
  (\frac{e^{-2\pi r_{horizon} \omega}}{1-e^{-2\pi  r_{horizon} \omega}})^{3}\times\nonumber \\&&(\frac{-2\pi \omega(r_{horizon}-r_{0,horizon})+\ln \frac{1-e^{-2\pi  r_{0,horizon} \omega}}{1-e^{-2\pi  r_{horizon} \omega}}}{\bar{\omega}})^{-\frac{15f-18}{5f+2}}
  \sim\nonumber \\&&
 (\frac{\Gamma_{0}^{3}}{36(4\pi)^{3}})^{\frac{1}{2}}(\frac{3^{11}(fA)^{15}(1-f)^{3}}{(2C)^{11}})^{\frac{1}{8}}
  (\frac{e^{-2\pi \frac{R}{\dot{R}} \omega}}{1-e^{-2\pi  \frac{R}{\dot{R}} \omega}})^{3}\times\nonumber \\&&(\frac{2\pi \omega(\frac{R_{0}}{\dot{R}_{0}}-\frac{R}{\dot{R}})+\ln \frac{1-e^{-2\pi  \frac{R_{0}}{\dot{R}_{0}} \omega}}{1-e^{-2\pi  \frac{R}{\dot{R}} \omega}}}{\bar{\omega}})^{-\frac{15f-18}{5f+2}}
\label{w40}
\end{eqnarray}
where \emph{k} is the co-moving wavenumber. With the wavenumber $k = k_{0} = 0.002Mpc^{-1}$, the
combined measurement from WMAP+BAO+SN of $\Delta_{R}^{2}$ is reported by the  WMAP7 data \cite{m20} as:
\begin{eqnarray}
\Delta_{R}^{2}=(2.455\pm 0.096)\times 10^{-19}\label{w41}
\end{eqnarray}
Using this equation and equation(\ref{w40}), and choosing (A=1, f=1/2, $\dot{R}=0.1$, $\omega=4.6 (GeV)$, $\Gamma_{0}$=1), we obtain the radial distance between our brane and another brane, $R=(1.5 GeV)^{-1}$. This result is consistent with previous calculations \cite{m21}.

Another important perturbation parameter is the spectral index $n_{s}$ which is given by:
\begin{eqnarray}
&&n_{s}-1=-\frac{d ln\Delta_{R}^{2}}{d lnk}=\nonumber \\&&\frac{15f-18}{8fA}(\frac{\ln <B>-\ln <B_{0}>}{\bar{\omega}})^{\frac{-8f}{5f+2}}=\nonumber \\&&
\frac{15f-18}{8fA}(\frac{\ln \frac{e^{-2\pi r_{horizon} \omega}}{1-e^{-2\pi  r_{horizon} \omega}}-\ln \frac{e^{-2\pi r_{0,horizon} \omega}}{1-e^{-2\pi  r_{0,horizon} \omega}}}{\bar{\omega}})^{\frac{-8f}{5f+2}}\sim\nonumber \\&&\frac{15f-18}{8fA}(\frac{-2\pi \omega(r_{horizon}-r_{0,horizon})+\ln \frac{1-e^{-2\pi  r_{0,horizon} \omega}}{1-e^{-2\pi  r_{horizon} \omega}}}{\bar{\omega}})^{\frac{-8f}{5f+2}}\sim\nonumber \\&&\frac{15f-18}{8fA}(\frac{2\pi \omega(\frac{R_{0}}{\dot{R}_{0}}-\frac{R}{\dot{R}})+\ln \frac{1-e^{-2\pi  \frac{R_{0}}{\dot{R}_{0}} \omega}}{1-e^{-2\pi  \frac{R}{\dot{R}} \omega}}}{\bar{\omega}})^{\frac{-8f}{5f+2}}
\label{w42}
\end{eqnarray}
where we have used  the thermal distribution in Eq.(\ref{w20}). In Fig. 2 we show the results for the spectral index in the intermediate
scenario as a function of $R^{-1}$, where \emph{R} is the orbital radial distance between tthe branes. In this plot we choose $R_{0}=0.45 (GeV)^{-1}$, $\omega=4.6 (GeV)$, $\dot{R}_{0}=0.01$, $\dot{R}=0.1$, \emph{A}=1 and $f=\frac{1}{2}$. As can be seen from  Fig.2, the spectral index decreases rapidly when the distance between the  branes  increases. By comparing Fig. 1 and Fig. 2, we find that the $N\simeq 50$ case leads to $n_{s}\simeq 0.96$. This result is compatible with observational data \cite{m5,m20,m22}. At this point, the radial distance between our brane and another brane is $R=(1.5 GeV)^{-1}$.
\begin{figure}\epsfysize=10cm
{ \epsfbox{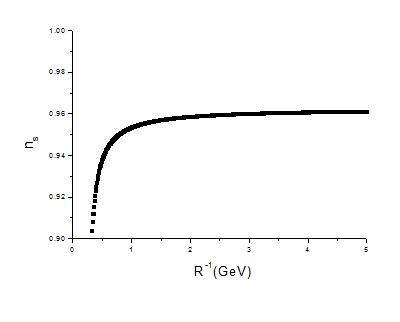}}\caption{The spectral index in intermediate
scenario as a function of $R^{-1}$ for $R_{0}=0.45 (GeV)^{-1}$, $\omega=4.6 (GeV)$, $\dot{R}_{0}=0.01$, $\dot{R}=0.1$, \emph{A}=1 and $f=\frac{1}{2}$. } \label{1}
\end{figure}

Using Eq.(\ref{w20}), we can calculate the tensor power spectrum and its spectral index as:
\begin{eqnarray}
&&\Delta_{T}^{2}=\frac{2H^{2}}{\pi^{2}}=\nonumber \\&&\frac{2(fA)^{2}}{\pi^{2}}(\frac{\ln <B>-\ln <B_{0}>}{\bar{\omega}})^{\frac{16(f-1)}{5f+2}}=\nonumber \\&&
\frac{2(fA)^{2}}{\pi^{2}}(\frac{\ln \frac{e^{-2\pi r_{horizon} \omega}}{1-e^{-2\pi  r_{horizon} \omega}}-\ln \frac{e^{-2\pi r_{0,horizon} \omega}}{1-e^{-2\pi  r_{0,horizon} \omega}}}{\bar{\omega}})^{\frac{16(f-1)}{5f+2}}\sim\nonumber \\&& \frac{2(fA)^{2}}{\pi^{2}}(\frac{-2\pi \omega(r_{horizon}-r_{0,horizon})+\ln \frac{1-e^{-2\pi  r_{0,horizon} \omega}}{1-e^{-2\pi  r_{horizon} \omega}}}{\bar{\omega}})^{\frac{16(f-1)}{5f+2}}\sim\nonumber \\&&\frac{2(fA)^{2}}{\pi^{2}}(\frac{2\pi \omega(\frac{R_{0}}{\dot{R}_{0}}-\frac{R}{\dot{R}})+\ln \frac{1-e^{-2\pi  \frac{R_{0}}{\dot{R}_{0}} \omega}}{1-e^{-2\pi  \frac{R}{\dot{R}} \omega}}}{\bar{\omega}})^{\frac{16(f-1)}{5f+2}}
\label{w43}
\end{eqnarray}
\begin{eqnarray}
&&n_{T}=-2\varepsilon=\nonumber \\&&-\frac{2-2f}{fA}(\frac{\ln <B>-\ln <B_{0}>}{\bar{\omega}})^{\frac{-8f}{5f+2}}=\nonumber \\&&
-\frac{2-2f}{fA}(\frac{\ln \frac{e^{-2\pi r_{horizon} \omega}}{1-e^{-2\pi  r_{horizon} \omega}}-\ln \frac{e^{-2\pi r_{0,horizon} \omega}}{1-e^{-2\pi  r_{0,horizon} \omega}}}{\bar{\omega}})^{\frac{-8f}{5f+2}}\sim\nonumber \\&&-\frac{2-2f}{fA}(\frac{-2\pi \omega(r_{horizon}-r_{0,horizon})+\ln \frac{1-e^{-2\pi  r_{0,horizon} \omega}}{1-e^{-2\pi  r_{horizon} \omega}}}{\bar{\omega}})^{\frac{-8f}{5f+2}}\sim\nonumber \\&&-\frac{2-2f}{fA}(\frac{2\pi \omega(\frac{R_{0}}{\dot{R}_{0}}-\frac{R}{\dot{R}})+\ln \frac{1-e^{-2\pi  \frac{R_{0}}{\dot{R}_{0}} \omega}}{1-e^{-2\pi  \frac{R}{\dot{R}} \omega}}}{\bar{\omega}})^{\frac{-8f}{5f+2}}
\label{w44}
\end{eqnarray}
These perturbations depend on the orbital radial distance between the branes. As we discussed before, these perturbations have a direct effect on the cosmic microwave background (CMB). Thus, we can observe the signature of interaction between the  branes by means of observational data.

 Another important parameter is tthe tensor-scalar ratio that has the following form:
\begin{eqnarray}
  &&R_{Tensor-scalar }=-(\frac{144(4\pi)^{3}(fA)^{4}}{\Gamma_{0}^{3}\pi^{4}T^{2}})^{\frac{1}{2}}\dot{H}H^{\frac{1}{2}}=\nonumber \\&&(\frac{144(4\pi)^{3}(fA)^{4}}{\Gamma_{0}^{3}\pi^{4}})^{\frac{1}{2}}(\frac{3^{11}(fA)^{15}(1-f)^{3}}{(2C)^{11}})^{\frac{1}{8}}
  \times\nonumber \\&&<B>^{3}(\frac{\ln <B>-\ln <B_{0}>}{\bar{\omega}})^{\frac{f+2}{5f+2}}=\nonumber \\&&
  (\frac{144(4\pi)^{3}(fA)^{4}}{\Gamma_{0}^{3}\pi^{4}})^{\frac{1}{2}}(\frac{3^{11}(fA)^{15}(1-f)^{3}}{(2C)^{11}})^{\frac{1}{8}}
  (\frac{e^{-2\pi r_{horizon} \omega}}{1-e^{-2\pi  r_{horizon} \omega}})^{3}\times\nonumber \\&&(\frac{\ln \frac{e^{-2\pi r_{horizon} \omega}}{1-e^{-2\pi  r_{horizon} \omega}}-\ln \frac{e^{-2\pi r_{0,horizon} \omega}}{1-e^{-2\pi  r_{0,horizon} \omega}}}{\bar{\omega}})^{\frac{f+2}{5f+2}}\sim\nonumber \\&&
 (\frac{144(4\pi)^{3}(fA)^{4}}{\Gamma_{0}^{3}\pi^{4}})^{\frac{1}{2}}(\frac{3^{11}(fA)^{15}(1-f)^{3}}{(2C)^{11}})^{\frac{1}{8}}
  (\frac{e^{-2\pi r_{horizon} \omega}}{1-e^{-2\pi  r_{horizon} \omega}})^{3}\times\nonumber \\&&(\frac{-2\pi \omega (r_{horizon}-r_{0,horizon})+\ln \frac{1-e^{-2\pi  r_{0,horizon} \omega}}{1-e^{-2\pi  r_{horizon} \omega}}}{\bar{\omega}})^{\frac{f+2}{5f+2}}
  \sim\nonumber \\&&
 (\frac{144(4\pi)^{3}(fA)^{4}}{\Gamma_{0}^{3}\pi^{4}})^{\frac{1}{2}}(\frac{3^{11}(fA)^{15}(1-f)^{3}}{(2C)^{11}})^{\frac{1}{8}}
  (\frac{e^{-2\pi \frac{R}{\dot{R}} \omega}}{1-e^{-2\pi  \frac{R}{\dot{R}} \omega}})^{3}\times\nonumber \\&&(\frac{2\pi \omega(\frac{R_{0}}{\dot{R}_{0}}-\frac{R}{\dot{R}})+\ln \frac{1-e^{-2\pi  \frac{R_{0}}{\dot{R}_{0}} \omega}}{1-e^{-2\pi  \frac{R}{\dot{R}} \omega}}}{\bar{\omega}})^{\frac{f+2}{5f+2}}
\label{w45}
\end{eqnarray}
In Fig. 3 we present the tensor-scalar ratio in the intermediate
scenario as a function of $R^{-1}$, where \emph{R} is the orbital radial distance between the branes. In this plot we choose $R_{0}=0.45 (GeV)^{-1}$, $\omega=4.6 (GeV)$, $\dot{R}_{0}=0.01$, $\dot{R}=0.1$, C=70, $\Gamma_{0}=1$, \emph{A}=1 and $f=\frac{1}{2}$. We observe that as the orbital radius distance between branes increases, the tensor-scalar ratio increases. By comparing Figs. 2 and 3, we notice that the standard case $n_{s} \simeq 0.96$, may be found in $0.01 < R_{Tensor-scalar } < 0.22$, which agrees with observational data \cite{m5,m20,m22}. At this stage, the radial  distance between our brane and another brane is $R=(1.5 GeV)^{-1}$.

\begin{figure}\epsfysize=10cm
{ \epsfbox{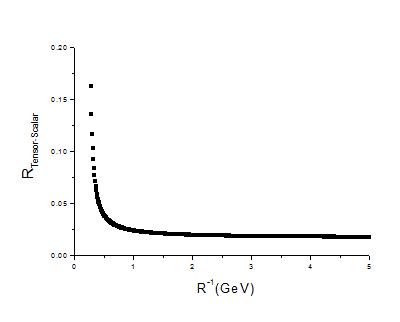}}\caption{The tensor-scalar ratio in intermediate
scenario as a function of $R^{-1}$ for $R_{0}=0.45 (GeV)^{-1}$, $\omega=4.6 (GeV)$, $\dot{R}_{0}=0.01$, $\dot{R}=0.1$, \emph{A}=1 and $f=\frac{1}{2}$. } \label{1}
\end{figure}

Now, we would like to consider the signature of interaction between branes in the context of logamediate inflation 
with scale factor:
\begin{eqnarray}
a(t)=a_{0}exp(A[\ln t]^{\lambda})\label{w46}
\end{eqnarray}
where \emph{A} is a constant parameter. This model is converted to power-law inflation for the $\lambda = 1$ case.
This scenario is applied for a number of scalar-tensor theories \cite{m13}. The effective potential of this solution is used in dark energy models \cite{m23}, supergravity,
Kaluza-Klein theories and super-string models \cite{m13,m24}. The number of e-folds in this case is given by \cite{m5}:
\begin{eqnarray}
N=\int_{t_{1}}^{t}H dt=A([\ln t]^{\lambda}-[\ln t_{1}]^{\lambda})
\label{w47}
\end{eqnarray}
where $t_{1}$ is the begining time of inflation. From Eqs. (\ref{w20}), (\ref{w24}), (\ref{w25}), (\ref{w26}), (\ref{w27}) and (\ref{w46}) we may find the inflaton \textbf{B} and also the orbital radial distance between the two branes:
\begin{eqnarray}
&&\ln <B>-\ln <B_{0}>=\tilde{\omega}\Xi(t)\rightarrow\nonumber \\&&
\ln \frac{e^{-2\pi r_{horizon} \omega}}{1-e^{-2\pi  r_{horizon} \omega}}-\ln \frac{e^{-2\pi r_{0,horizon} \omega}}{1-e^{-2\pi  r_{0,horizon} \omega}}=\tilde{\omega}\Xi(t)\rightarrow\nonumber \\&&-2\pi \omega (r_{horizon}-r_{0,horizon})+\ln \frac{1-e^{-2\pi  r_{0,horizon} \omega}}{1-e^{-2\pi  r_{horizon} \omega}}\sim\tilde{\omega}\Xi(t)\rightarrow\nonumber \\&&(2\pi \omega(\frac{R_{0}}{\dot{R}_{0}}-\frac{R}{\dot{R}})+\ln \frac{1-e^{-2\pi  \frac{R_{0}}{\dot{R}_{0}} \omega}}{1-e^{-2\pi  \frac{R}{\dot{R}} \omega}})\sim\tilde{\omega}\Xi(t)
\label{w48}
\end{eqnarray}
where $\tilde{\omega}=(\frac{6}{\Gamma_{0}}(\frac{2C}{3})^{\frac{3}{4}})^{\frac{1}{2}}((-4)^{5\lambda+3}(\lambda A)^{5})^{\frac{1}{8}}$ and $\Xi(t)=\gamma[\frac{5\lambda+3}{8},\frac{\ln t}{4}]$ ($\gamma$  is the incomplete gamma
function \cite{m25}). The potential in terms of the orbital radial distance between the two branes is presented as:
\begin{eqnarray}
&&V=\frac{2\lambda^{2}A^{2}[\ln(\Xi^{-1}(\frac{\ln <B>-\ln <B_{0}>}{\tilde{\omega}}))]^{2\lambda-2}}{(\Xi^{-1}(\frac{\ln <B>-\ln <B_{0}>}{\tilde{\omega}}))^{2}}=\nonumber \\&&\frac{2\lambda^{2}A^{2}[\ln(\Xi^{-1}(\frac{(2\pi \omega(\frac{R_{0}}{\dot{R}_{0}}-\frac{R}{\dot{R}})+\ln \frac{1-e^{-2\pi  \frac{R_{0}}{\dot{R}_{0}} \omega}}{1-e^{-2\pi  \frac{R}{\dot{R}} \omega}})}{\tilde{\omega}}))]^{2\lambda-2}}{(\Xi^{-1}(\frac{(2\pi \omega(\frac{R_{0}}{\dot{R}_{0}}-\frac{R}{\dot{R}})+\ln \frac{1-e^{-2\pi  \frac{R_{0}}{\dot{R}_{0}} \omega}}{1-e^{-2\pi  \frac{R}{\dot{R}} \omega}})}{\tilde{\omega}}))^{2}}
\label{w49}
\end{eqnarray}
This equation shows that the inflatonic potential on our brane depends on the orbital radial distance and the interaction potential between the two branes. In fact, the interaction between branes causes  inflation of our universe.

Now, we obtain the slow-roll parameters of the model in this case:
\begin{eqnarray}
&&\epsilon=\frac{[\ln(\Xi^{-1}(\frac{(2\pi \omega(\frac{R_{0}}{\dot{R}_{0}}-\frac{R}{\dot{R}})+\ln \frac{1-e^{-2\pi  \frac{R_{0}}{\dot{R}_{0}} \omega}}{1-e^{-2\pi  \frac{R}{\dot{R}} \omega}})}{\tilde{\omega}}))]^{1-\lambda}}{\lambda A}\nonumber \\&&\eta=\frac{2[\ln(\Xi^{-1}(\frac{(2\pi \omega(\frac{R_{0}}{\dot{R}_{0}}-\frac{R}{\dot{R}})+\ln \frac{1-e^{-2\pi  \frac{R_{0}}{\dot{R}_{0}} \omega}}{1-e^{-2\pi  \frac{R}{\dot{R}} \omega}})}{\tilde{\omega}}))]^{1-\lambda}}{\lambda A}
\label{w50}
\end{eqnarray}
In this case, like the intermediate case, as the distance between the  branes decreases, more inflatons are created in the brane-antibrane system, the slow-roll parameters increase, and the universe inflates.

Using Eqs (\ref{w20}), (\ref{w47}) and (\ref{w48}), the number of e-folds between the  two fields $B_{1}$ and \textbf{B}(t) can be obtained as:
\begin{eqnarray}
&&N=A([\ln(\Xi^{-1}(\frac{(2\pi \omega(\frac{R_{0}}{\dot{R}_{0}}-\frac{R}{\dot{R}})+\ln \frac{1-e^{-2\pi  \frac{R_{0}}{\dot{R}_{0}} \omega}}{1-e^{-2\pi  \frac{R}{\dot{R}} \omega}})}{\tilde{\omega}}))]^{\lambda}\nonumber \\&&-[\ln(\Xi^{-1}(\frac{(2\pi \omega(\frac{R_{0}}{\dot{R}_{0}}-\frac{R_{1}}{\dot{R}_{1}})+\ln \frac{1-e^{-2\pi  \frac{R_{0}}{\dot{R}_{0}} \omega}}{1-e^{-2\pi  \frac{R_{1}}{\dot{R}_{1}} \omega}})}{\tilde{\omega}}))]^{\lambda})=\nonumber \\&&A([\ln(\Xi^{-1}(\frac{(2\pi \omega(\frac{R_{0}}{\dot{R}_{0}}-\frac{R}{\dot{R}})+\ln \frac{1-e^{-2\pi  \frac{R_{0}}{\dot{R}_{0}} \omega}}{1-e^{-2\pi  \frac{R}{\dot{R}} \omega}})}{\tilde{\omega}}))]^{\lambda}-[\lambda A]^{\frac{\lambda}{1-\lambda}})
\label{w51}
\end{eqnarray}
where $R_{1}$ is the the orbital radial distance between the two branes at the begining of the inflationary epoch when ($\epsilon=1$). Using the above equation, the orbital radial distance between the two branes in the
inflationary period could be obtained in terms of the number of e-folds as:
\begin{eqnarray}
R-R_{0}=\frac{\tilde{\omega}\dot{R}^{2}\dot{R}^{2}_{0}}{2\pi \omega}\Xi[exp(\frac{N}{A}+(\lambda A)^{\frac{\lambda}{1-\lambda}})^{\frac{1}{\lambda}}]
\label{w52}
\end{eqnarray}
This equation shows that, in this case, like the intermediate case, the number of e-fields depends on the orbital radial distance between the branes. This is because as  the distance between the branes decreases, the number of inflatons, which has direct effects on the number of e-folds, increases.

Also, the scalar and tensor power spectrum in this case are given by:
\begin{eqnarray}
&&\Delta_{R}^{2}=(\frac{\Gamma_{0}^{3}}{36(4\pi)^{3}})^{\frac{1}{2}}(\frac{3^{11}(\lambda A)^{15}}{(2C)^{11}})^{\frac{1}{8}}(\frac{e^{-2\pi r_{horizon} \omega}}{1-e^{-2\pi  r_{horizon} \omega}})^{-3}\nonumber \\&&\times exp(-\frac{15}{8}[\ln(\Xi^{-1}(\frac{(2\pi \omega(\frac{R_{0}}{\dot{R}_{0}}-\frac{R}{\dot{R}})+\ln \frac{1-e^{-2\pi  \frac{R_{0}}{\dot{R}_{0}} \omega}}{1-e^{-2\pi  \frac{R}{\dot{R}} \omega}})}{\tilde{\omega}}))])\nonumber \\&&\times
[[\ln(\Xi^{-1}(\frac{(2\pi \omega(\frac{R_{0}}{\dot{R}_{0}}-\frac{R}{\dot{R}})+\ln \frac{1-e^{-2\pi  \frac{R_{0}}{\dot{R}_{0}} \omega}}{1-e^{-2\pi  \frac{R}{\dot{R}} \omega}})}{\tilde{\omega}}))]^{\lambda}]^{15\frac{\lambda-1}{8\lambda}}
\nonumber \\&&\times exp(-3\tilde{\omega}\Xi(exp(\ln(\Xi^{-1}(\frac{(2\pi \omega(\frac{R_{0}}{\dot{R}_{0}}-\frac{R}{\dot{R}})+\ln \frac{1-e^{-2\pi  \frac{R_{0}}{\dot{R}_{0}} \omega}}{1-e^{-2\pi  \frac{R}{\dot{R}} \omega}})}{\tilde{\omega}})))))
\label{w53}
\end{eqnarray}
\begin{eqnarray}
\Delta_{T}^{2}=\frac{2\lambda^{2}A^{2}[\ln(\Xi^{-1}(\frac{(2\pi \omega(\frac{R_{0}}{\dot{R}_{0}}-\frac{R}{\dot{R}})+\ln \frac{1-e^{-2\pi  \frac{R_{0}}{\dot{R}_{0}} \omega}}{1-e^{-2\pi  \frac{R}{\dot{R}} \omega}})}{\tilde{\omega}}))]^{2-2\lambda}}{\pi^{2}(\Xi^{-1}(\frac{(2\pi \omega(\frac{R_{0}}{\dot{R}_{0}}-\frac{R}{\dot{R}})+\ln \frac{1-e^{-2\pi  \frac{R_{0}}{\dot{R}_{0}} \omega}}{1-e^{-2\pi  \frac{R}{\dot{R}} \omega}})}{\tilde{\omega}}))^{2}}
\label{w54}
\end{eqnarray}
These spectra in the context of logamediate inflation, like intermediate inflation, change with an increase or decrease in the orbital radial distance between the branes. The spectral index in this case has the following forms:
\begin{eqnarray}
&&n_{s}-1=-\frac{15(\lambda-1)}{8\lambda A}[[\ln(\Xi^{-1}(\frac{(2\pi \omega(\frac{R_{0}}{\dot{R}_{0}}-\frac{R}{\dot{R}})+\ln \frac{1-e^{-2\pi  \frac{R_{0}}{\dot{R}_{0}} \omega}}{1-e^{-2\pi  \frac{R}{\dot{R}} \omega}})}{\tilde{\omega}}))]^{\lambda}]^{-1}\nonumber \\&&n_{T}=-\frac{2[\ln(\Xi^{-1}(\frac{(2\pi \omega(\frac{R_{0}}{\dot{R}_{0}}-\frac{R}{\dot{R}})+\ln \frac{1-e^{-2\pi  \frac{R_{0}}{\dot{R}_{0}} \omega}}{1-e^{-2\pi  \frac{R}{\dot{R}} \omega}})}{\tilde{\omega}}))]^{1-\lambda}}{\lambda A}
\label{w55}
\end{eqnarray}
\begin{figure}\epsfysize=10cm
{ \epsfbox{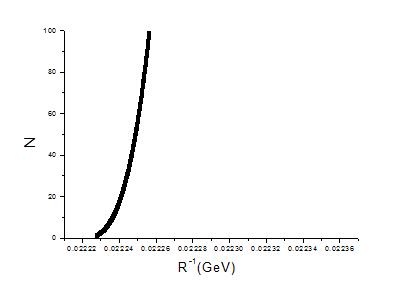}}\caption{The number of e-folds \emph{N} in logamediate inflation
scenario as a function of $R^{-1}$ for $R_{0}=0.45 (GeV)^{-1}$, $\omega=4.6 (GeV)$, $\dot{R}_{0}=0.01$, $\dot{R}=0.1$, $\lambda=10$, \emph{A}=1 and $f=\frac{1}{2}$. } \label{1}
\end{figure}
\begin{figure}\epsfysize=10cm
{ \epsfbox{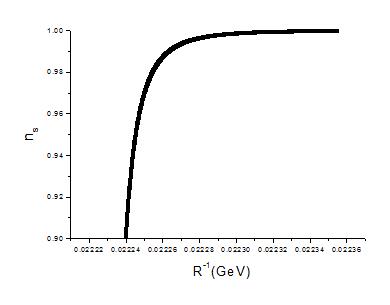}}\caption{The spectral index in logamediate inflation
scenario as a function of $R^{-1}$ for $R_{0}=0.45 (GeV)^{-1}$, $\omega=4.6 (GeV)$, $\dot{R}_{0}=0.01$, $\dot{R}=0.1$, $\lambda=10$, \emph{A}=1 and $f=\frac{1}{2}$. } \label{1}
\end{figure}
In Figs. 4 and 5, we present the number of e-folds \emph{N} and the spectral index for the  logamediate inflation
scenario as a function of $R^{-1}$, where R is the orbital radial distance between the  branes. In these plots, we choose $R_{0}=0.45 (GeV)^{-1}$, $\omega=4.6 (GeV)$, $\dot{R}_{0}=0.01$, $\dot{R}=0.1$, $\lambda=10$, \emph{A}=1 and $f=\frac{1}{2}$. In this case, like the intermediate case, we find that the number of e-folds \emph{N} and the spectral index are much larger for smaller orbital radial distance between branes. This is because, as the distance between the branes becomes smaller, the temperature becomes larger, and the thermal radiation of the inflatons  enhances.

Finally, we could find the tensor-scalar ratio in terms of the orbital radial distance between two branes:
\begin{eqnarray}
&&R_{Tensor-scalar }=(\frac{144(4\pi)^{3}}{\Gamma_{0}^{3}\pi^{4}})^{\frac{1}{2}}(\frac{(2C)^{11}(\lambda A)}{3^{11}})^{\frac{1}{8}}(\frac{e^{-2\pi r_{0,horizon} \omega}}{1-e^{-2\pi  r_{0,horizon} \omega}})^{3}\nonumber \\&&\times exp(-\frac{1}{8}[\ln(\Xi^{-1}(\frac{(2\pi \omega(\frac{R_{0}}{\dot{R}_{0}}-\frac{R}{\dot{R}})+\ln \frac{1-e^{-2\pi  \frac{R_{0}}{\dot{R}_{0}} \omega}}{1-e^{-2\pi  \frac{R}{\dot{R}} \omega}})}{\tilde{\omega}}))])\nonumber \\&&\times
[[\ln(\Xi^{-1}(\frac{(2\pi \omega(\frac{R_{0}}{\dot{R}_{0}}-\frac{R}{\dot{R}})+\ln \frac{1-e^{-2\pi  \frac{R_{0}}{\dot{R}_{0}} \omega}}{1-e^{-2\pi  \frac{R}{\dot{R}} \omega}})}{\tilde{\omega}}))]^{\lambda}]^{\frac{-31(\lambda-1)}{8\lambda}}
\nonumber \\&&\times exp(3\tilde{\omega}\Xi [exp(\ln(\Xi^{-1}(\frac{(2\pi \omega(\frac{R_{0}}{\dot{R}_{0}}-\frac{R}{\dot{R}})+\ln \frac{1-e^{-2\pi  \frac{R_{0}}{\dot{R}_{0}} \omega}}{1-e^{-2\pi  \frac{R}{\dot{R}} \omega}})}{\tilde{\omega}})))])
\label{w56}
\end{eqnarray}
In Fig. 6 we present the tensor-scalar ratio in the logamediate
scenario as a function of $R^{-1}$, where \emph{R} is the orbital radial distance between the branes. In this plot we choose $R_{0}=0.45 (GeV)^{-1}$, $\omega=4.6 (GeV)$, $\dot{R}_{0}=0.01$, $\dot{R}=0.1$, C=70, $\Gamma_{0}=1$, $\lambda=10$, \emph{A}=1 and $f=\frac{1}{2}$. In this case, like the intermediate case, with an  increase in  the orbital radial distance between branes, the tensor-scalar ratio increases. By comparing Figs. 5 and 6, we notice that the standard case $n_{s} \simeq 0.96$, may be found in $0.01 < R_{tensor-scalar } < 0.22$, which agrees with observational data \cite{m5,m20,m22}. At this stage, the radial distance between our brane and another brane is $R=(0.02225 GeV)^{-1}$.
\begin{figure}\epsfysize=10cm
{ \epsfbox{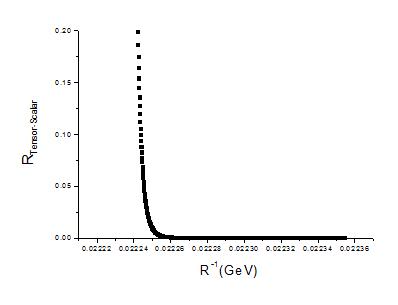}}\caption{The tensor-scalar ratio in logamediate inflation
scenario as a function of $R^{-1}$ for $R_{0}=0.45 (GeV)^{-1}$, $\omega=4.6 (GeV)$, $\dot{R}_{0}=0.01$, $\dot{R}=0.1$, $\lambda=10$, \emph{A}=1 and $f=\frac{1}{2}$. } \label{1}
\end{figure}

\section{Summary and Discussion} \label{sum}
In this research,  we calculate the thermal distribution of inflatons near the apparent  horizon in a brane-antibrane system, and show that the energy density, slow-roll, number of e-folds and
perturbation parameters can be given  in terms of  the orbital radius of  the brane motion in extra dimensions. According to our results, when the distance between branes  increases, the number of e-folds and the spectral index for both intermediate and logamediate
models decrease rapidly; however the tensor-scalar ratio increases. This is because, as the  separate distance between branes decreases, the interaction potential increases, and at higher energies, there exist more channels for inflaton production near the apparent 
horizon in the brane-antibrane system; consequently, the effect of inflaton radiation from this horizon on cosmic inflation becomes systematically
more effective. We find that the $N\simeq 50$ case leads to $n_{s}\simeq 0.96$. This standard case may be found in $0.01 < R_{tensor-scalar } < 0.22$, which agrees with observational data \cite{m5,m20,m22} (We note some new observational data has been obtained, but we believe that our models will fit this as well. This work in under progress). At this point, the radial distance between our brane and another brane is $R=(1.5 GeV)^{-1}$ in the intermediate model and $R=(0.02225 GeV)^{-1}$ in the logamediate model. 

There is no underlying data that is necessary for this work, and there is no special funding.

The author declares that there is no conflict of interest regarding the publication of this paper

\end{document}